\documentclass[12pt]{article}

\let\section=\subsection     \let\subsection=\subsubsection                
\usepackage{graphicx}
\usepackage{amsmath} 
\usepackage{epsfig} 

\begin{document}

\begin{center}
   {\large \bf INSTANTONS AND GLUEBALLS}\\[5mm]
   H. FORKEL \\[5mm]
   {\small \it  Institut f\"{u}r Theoretische Physik, Universit\"{a}t
   Heidelberg,\\ Philosophenweg 19, D-69120 Heidelberg, Germany  \\[8mm] } 
\end{center}

\begin{abstract}\noindent
We investigate the impact of instantons on scalar glueball properties in a 
largely model-independent analytical approach based on the  instanton-improved
operator product expansion (IOPE) of the $0^{++}$  glueball correlator. The
instanton contributions turn out to be dominant, to substantially improve the
consistency of the correponding QCD sum rules, and to increase the
glueball residue about fivefold.  \end{abstract}

\section{Introduction}

As the hadrons with the largest gluon content (despite probably 
sizeable quarkonium admixtures), glueballs provide a privileged
source of  information on how glue manifests itself in hadronic 
bound states (in their rest frame).  Progress in understanding 
glueball structure may therefore even help to unravel the elusive 
gluonic component of the classical hadrons.    

The intricacies of hadronic glue are exemplified by the problems encountered 
when implementing various effective, gluonic degrees of freedom, such  as
constituent gluons, strings, or flux tubes, into hadron models. The
resulting effects are generally less prominent and more ambiguous than
those of the quark substructure, for which phenomenologically  rather
successful effective degrees of freedom are provided, e.g., by constituent
quarks. The problems in modeling the gluon  sector are partially
related to the fact that gluons, unlike quarks, do not  carry internal, global
quantum numbers (like flavor or charge) to which  external probes can couple.
Unambiguous and transparent gluonic effects in  hadrons are therefore much
harder to identify.

Model-independent and more directly QCD-based approaches are thus called 
upon to clarify the structure of hadronic glue and to identify potentially 
dominant gluon field configurations. Promising candidates for the latter, 
especially  in the $0^{++}$ glueball channel, are instantons \cite{sch98} (i.e. 
the strong, coherent gluon fields which, by mediating tunneling processes, 
give rise to the $\theta $-angle of the QCD vacuum), as qualitative arguments
\cite  {nov280,shu82,for00} and instanton-liquid model results \cite{sch95} 
suggest. In the following,  we will give a brief summary of work in an 
analytical approach to the scalar glueball correlator at short distances  
\cite{for00,for01}, based on an instanton-improved operator product  expansion
(IOPE) and QCD sum-rule techniques, in which these issues can  be
addressed, and glueball properties calculated, in a largely  model-independent
fashion.  

\section{IOPE sum rules}

Consider the correlation function 
\begin{equation}
\Pi \left( -q^{2}\right) =i\int d^{4}xe^{iqx}\left\langle 0|T\,O_{S}\left(
x\right) O_{S}\left( 0\right) |0\right\rangle   \label{corr}
\end{equation}
where the interpolating field 
\begin{equation}
O_{S}=\alpha _{s}G_{\mu \nu }^{a}G^{a,\mu \nu }  \label{intpol}
\end{equation}
carries the quantum numbers of the scalar ($0^{++}$) glueball. The standard
OPE separates this correlator into contributions from hard field modes (with
typical momenta $k\geq \mu \sim 1/2$ GeV), contained in perturbative Wilson
coefficients, and from soft ($k<\mu $) modes contained in the vacuum
expectation values of composite QCD operators, the so-called condensates.
The perturbative Wilson coefficients of the low-dimensional operators can be
found to $O\left( \alpha _{s}\right) $ in \cite{bag90,nar98}.  

Remarkably, the nonperturbative power (i.e. condensates) corrections
to this conventional OPE turn out to be small and are, except for the
lowest-dimensional gluon condensate contribution, negligible. The bulk of
the nonperturbative physics reponsible for the strong binding in the scalar
glueball channel should therefore manifest itself in the Wilson
coefficients, i.e. predominantly via direct instantons \cite{nov280,shu82}
which are neglected in the standard analyses \cite{bag90,nar98}. This
expectation is strengthened by the fact that instantons couple
particularly strongly to the gluonic $0^{++}$ interpolators. 

In Ref. \cite{for00,for01}, we have evaluated the direct-instanton
contributions and implemented them into the corresponding IOPE sum rules,
which are based on the Borel-transformed moments 
\begin{equation}
\mathcal{R}_{k}\left( \tau \right) =\hat{B}\left[ \left( -Q^{2}\right)
^{k}\Pi \left( Q^{2}\right) \right]   \label{momdef}
\end{equation}
(for the explicit form of the Borel operator $\hat{B}$ see, 
e.g. 
\cite{bag90}) with $k \in \left\{ -1,0,1,2 \right\}$. 
Besides the standard OPE
contributions $\mathcal{R}_{k}^{\left( OPE\right) }$ of 
Ref. \cite{bag90},
the IOPE includes the instanton contributions \cite{for00} 
\begin{eqnarray}
\mathcal{R}_{k}^{\left( I+\bar{I}\right) }\left( \tau \right)  
&=&-2^{6}\pi^{2}\bar{n}\left( \frac{-\partial }{\partial \tau }
\right)^{k+1} \nonumber \\ &&\times \left\{ \xi^{2}e^{-\xi}
\left[ \left(1+\xi\right) K_{0}\left( \xi\right) +\left(2+
\xi+\frac{2}{\xi}\right) K_{1} \left( \xi\right) \right] -2\right\},
\label{ri}  \end{eqnarray}
($k\geq -1$, $\xi\equiv \bar{\rho}^{2}/\left( 2\tau \right) $). The
semiclassical result (\ref{ri}) neglects $O\left( \hbar \right) $
corrections (which are suppressed by the large instanton action $S_{I}\left( 
\bar{\rho}\right) \sim 10\hbar $) and multi-instanton correlations (since
$\left| x\right| \sim \left| Q^{-1}\right| \leq 0.2$ fm is small compared
with the average instanton separation $\bar{R}\sim 1$ fm). We have also
removed the soft subtraction term $-\Pi ^{\left( I+\bar{I}\right) }\left(
0\right) =-2^{7}\pi ^{2}\bar{n}$ \ to avoid double-counting with the
condensates. Below, the average instanton size $\bar{\rho}\simeq (1/3)$ fm
and density $\bar{n}\simeq (1/2)$ fm$^{-4}$ will be fixed at the values
obtained (approximately) from instanton vacuum model \cite{sch98} and
lattice \cite{instlat} simulations. 

To obtain sum rules, the IOPE expressions are matched to their
``phenomenological'' counterparts, which are derived from the borelized
dispersive representation 
\begin{equation}
\mathcal{R}_{k}^{\left( ph\right) }\left( \tau \right) =\frac{1}{\pi }  
\int_{0}^{\infty }dss^{k} {\rm Im}\Pi ^{\left( ph\right) }\left( s\right)
e^{-s\tau }+\delta _{k,-1}\Pi ^{\left( ph\right) }\left( 0\right) 
\end{equation}
where the spectral function ${\rm Im}\Pi ^{\left( ph\right) }\left(
s\right) $ contains a glueball pole contribution and an effective continuum
from the dispersive cut of the IOPE, starting at an effective threshold
$s_{0}$. This corresponds to 
\begin{equation}
{\rm Im}\Pi ^{\left( ph\right) }\left( s\right) =\pi
f_{G}^{2}m_{G}^{4}\delta \left( s-m_{G}^{2}\right) +{\rm Im}\left[ \Pi
^{\left( OPE\right) }+\Pi ^{\left( I+\bar{I}\right) }\right] \left( s\right)
\theta \left( s-s_{0}\right) .  \label{specdens}
\end{equation}
The instanton continuum contributions 
\begin{equation}
{\rm Im}\Pi ^{\left( I+\bar{I}\right) }\left( s\right) =-2^{4}\pi ^{4}\bar{n%
}\bar{\rho}^{4}s^{2}J_{2}\left( \sqrt{s}\bar{\rho}\right) Y_{2}\left( \sqrt{s%
}\bar{\rho}\right)   \label{icont}
\end{equation}
($J_{2}$ ($Y_{2}$) are Bessel (Neumann) functions), introduced in \cite
{for00}, will play an essential role in the subsequent analysis. The IOPE
sum rules can then be written as 
\begin{equation}
\mathcal{R}_{k}^{\left( IOPE\right) }\left( \tau \right) \equiv \mathcal{R}%
_{k}^{\left( I+\bar{I}\right) }\left( \tau \right) +\mathcal{R}_{k}^{\left(
OPE\right) }\left( \tau \right) =\mathcal{R}_{k}^{\left( ph\right) }\left(
\tau ,s_{0}\right) .  \label{iopesrs}
\end{equation}

The subtraction constant $\Pi ^{\left( ph\right) }\left( 0\right) $ in\ the
$\mathcal{R}_{-1}$ sum rule (regularized by removing the high-momentum
contributions) is related to the gluon condensate by the low-energy theorem 
\cite{nov280} 
\begin{equation}
\Pi \left( 0\right) =\frac{32\pi }{b_{0}}\left\langle \alpha
G^{2}\right\rangle .  \label{let}
\end{equation}
This relation provides an important consistency check for the sum-rule
analysis, as discussed below.

\section{Results and conclusions}

The quantitative analysis of the sum rules (\ref{iopesrs}) amounts to
determining those values of the glueball  parameters and $s_0$ in Eq.
(\ref{specdens}) for which both sides optimally match in the fiducial $\tau $
domain (determined such that the approximations on both sides of the sum
rules are expected to be reliable \cite{for00}). The previously neglected
instanton contributions turn out to be dominant and render (\ref{iopesrs}) 
the first overall consistent set of QCD sum rules in the scalar glueball
channel.

In particular, the IOPE resolves two long-standing flaws of the
earlier sum rules \cite{nov280,bag90,nar98}: the mutual inconsistency between
different Borel moments and the inconsistency with the low-energy theorem
(\ref{let}). Even the previously deficient and usually discarded
lowest-moment ($\mathcal{R}_{-1}$) sum rule is rendered consistent both
with the higher-moment sum rules and the low-energy theorem. (A subsequent
analysis of the related gaussian variant of these sum rules can be found in
Ref. \cite{har00}.) No evidence for a low-lying ($m\ll 1$ GeV) gluonium state
(or any state strongly coupled to gluonic interpolators), which had been
argued for on the basis of this sum rule \cite{nov280}, remains. 
        
All four IOPE sum rules show an unprecedented degree of consistency and
allow for a simultaneous 3-parameter fit to the glueball mass, its coupling,
and the continuum threshold $s_{0}$. In Fig. 1 the glueball pole
contribution (solid line) to the optimized $\mathcal{R}_{0}$ sum rule is 
compared with the remaining components (OPE with subtracted OPE-continuum
(dash-double-dotted), instanton contribution (dashed), instanton continuum
(dash-dotted), and their sum (dotted)). Both parts match almost perfectly over
the whole fiducial region, mostly due to the dominant instanton
contributions. The remaining sum rules (including the $\mathcal{R}_{-1}$ sum
rule) show a similarly high degree of stability.  Together with the mutual
agreement of all four IOPE sum rules (in the typical range of uncertainty)
and their consistency with the low-energy theorem, this indicates that the
IOPE provides a sufficiently complete description of the short-distance
glueball correlator. The most dramatic quantitative effect of the direct
instantons is to increase the residuum of the glueball pole, $f_{G}^{2}$, by
about a factor of five. From the $\mathcal{R}_{2}$ sum rule (likely to be the
most reliable one since it receives the strongest relative pole contribution)
we obtain $m_{G}=1.53\pm 0.2$ GeV and $f_{G}=1.01\pm 0.25$ GeV.

\begin{figure}[htb]                                  
\begin{center}
\vskip -7mm                          
\rotatebox{270}{\scalebox{0.31}{\includegraphics{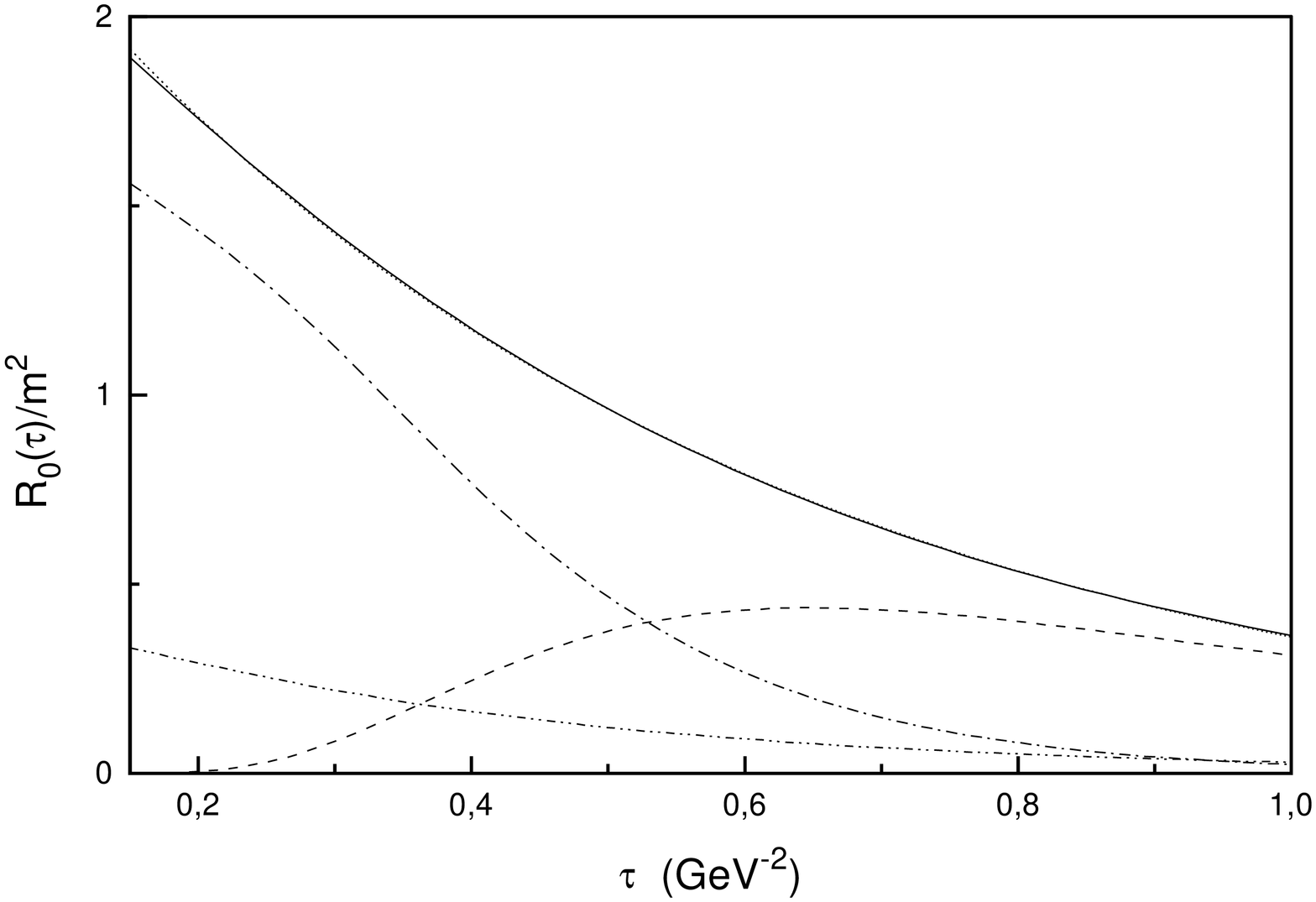}}}
\caption{Contributions to the $\mathcal{R}_{0}$ sum rule, as explained in the text.}       
\label{fig1}                             
\end{center}                           
\end{figure}
The instanton continuum contributions, Eq. (\ref{icont}), are
indispensable for the overall consistency and stability of the sum rules and
shed new light on the spectral content of the glueball correlator. For once,
they compensate, together with  the perturbative terms, the pole contribution
and thereby lead to an improved description of the correlator towards low
momenta. This reconciles the sum rules with the low-energy theorem in the
$Q\rightarrow 0$ limit. In the opposite limit, i.e. for $\tau \rightarrow 0$,
the instanton part of the continuum remains effective despite the suppression
incurred by funnelling a sizeable momentum through a coherent 
field. This indicates that small-instanton physics accounts for part of
the higher-lying strength in (\ref{corr}) and may play a rather prominent
role in excited glueball states. 

In contrast to previously studied IOPE sum rules \cite{shu83}, built on  
quark-based correlators, those for the scalar glueball are the first
where i) the instanton contributions do not enter via topological quark
zero-modes and where ii) the sum rules reach a satisfactory (though not
excellent) level of consistency even without any perturbative and soft
contributions, i.e. with the instanton terms alone. The latter result
explains why instanton models find scalar glueball properties similar to
those obtained above \cite{sch95}. It also illustrates that the
quantitative impact of the instanton contributions can be judged only if all
contributions are consistently taken into account. At present, the IOPE seems
to be the only controlled and analytical framework in which this is possible. 

The predominance and approximate self-sufficiency of the instanton
contributions indicates that instantons may generate the bulk of the forces
which bind the scalar glueball. A further, striking consequence of the
exceptionally strong instanton contributions is that the scales of the
predicted $0^{++}$ glueball properties are approximately set by the bulk
features of the instanton size distribution. More specifically, neglecting
the standard OPE contributions, one finds  
\begin{eqnarray}
m_{G} &\sim &\bar{\rho}^{-1}, \\
f_{G}^{2} &\sim &\bar{n}\bar{\rho}^{2}.
\end{eqnarray}
Conceptually, the main virtue of these scaling relations lies in
establishing an explicit link between fundamental vacuum and hadron
properties. They could also be of practical use, e.g. for the test of
instanton vacuum models, to provide constraints for glueball model building,
or generalized to finite temperature and baryon density, where the instanton
distribution  changes. If the glueball size $r_{G}$ scales like its Compton
wavelength, one furthermore has $r_{G}\sim \bar{\rho}$, in agreement with the
lattice calculations \cite{def92} which find $r_{G} \sim 0.2$~fm. 

It is a pleasure to thank the organizers for this interesting workshop and 
the Deutsche Forschungsgemeinschaft for financial support under Habilitation
Grant Fo 156/2-1.

\end{document}